\documentclass[sigconf,final]{acmart}
\AtBeginDocument{%
  }

\usepackage{graphicx}
\usepackage{subcaption} 

\usepackage{booktabs}
\usepackage{multirow}
\usepackage{amsmath}
\usepackage{setspace}
\usepackage{tabularx}
\usepackage{graphicx}

\usepackage{xcolor}

\setcopyright{acmlicensed}
\copyrightyear{2018}
\acmYear{2018}
\acmDOI{XXXXXXX.XXXXXXX}
\acmConference[Conference acronym 'XX]{Make sure to enter the correct
  conference title from your rights confirmation email}{June 03--05,
  2018}{Woodstock, NY}
\acmISBN{978-1-4503-XXXX-X/2018/06}

\begin{document}

\title{CardRewriter: Leveraging Knowledge Cards for Long-Tail Query Rewriting on Short-Video Platforms}


\author{Peiyuan Gong$^\dagger$}
\affiliation{%
  \institution{GSAI, Renmin University of China}
  \city{Beijing}
  \country{China}}
\email{pygongnlp@gmail.com}

\author{Feiran Zhu$^\dagger$}
\author{Yaqi Yin$^\dagger$}
\affiliation{%
  \institution{Kuaishou Technology}
  \city{Hangzhou, Beijing}
  \country{China}}
\email{{zhufeiran03,yinyaqi}@kuaishou.com} 

\author{Chenglei Dai}
\author{Chao Zhang}
\affiliation{%
  \institution{Kuaishou Technology}
  \city{Hangzhou, Beijing}
  \country{China}}
\email{{zhangchao,daichenglei}@kuaishou.com}

\thanks{$^\dagger$ Equal Contribution. Work done when Peiyuan Gong was an intern at Kuaishou.}
\thanks{$^*$ Corresponding author.}

\author{Kai Zheng}
\author{Wentian Bao}
\affiliation{%
  \institution{unaffiliated}
  \city{Beijing}
  \country{China}}
\email{zhengk92@gmail.com}
\email{wb2328@columbia.edu}

\author{Jiaxin Mao$^*$}
\affiliation{%
  \institution{GSAI, Renmin University of China}
  \city{Beijing}
  \country{China}}
\email{maojiaxin@gmail.com}

\author{Yi Zhang$^*$}
\affiliation{%
  \institution{Kuaishou Technology}
  \city{Beijing}
  \country{China}}
\email{zhangyi49@kuaishou.com}

\renewcommand{\shortauthors}{Gong et al.}

\begin{abstract}
Short-video platforms have rapidly become a new generation of information retrieval systems, where users formulate queries to access desired videos. However, user queries, especially long-tail ones, often suffer from spelling errors, incomplete phrasing, and ambiguous intent, resulting in mismatches between user expectations and retrieved results.
While large language models (LLMs) have shown success in long-tail query rewriting within e-commerce, they struggle on short-video platforms, where proprietary content such as short videos, live streams, micro dramas, and user social networks falls outside their training distribution.
To address this challenge, we introduce \textbf{CardRewriter}, an LLM-based framework that incorporates domain-specific knowledge to enhance long-tail query rewriting. 
For each query, our method aggregates multi-source knowledge relevant to the query and summarizes it into an informative and query-relevant knowledge card. 
This card then guides the LLM to better capture user intent and produce more effective query rewrites.
We optimize CardRewriter using a two-stage training pipeline: supervised fine-tuning followed by group relative policy optimization, with a tailored reward system balancing query relevance and retrieval effectiveness. 
Offline experiments show that CardRewriter substantially improves rewriting quality for queries targeting proprietary content. 
Online A/B testing further confirms significant gains in long-view rate (LVR) and click-through rate (CTR), along with a notable reduction in initiative query reformulation rate (IQRR).
Since September 2025, CardRewriter has been deployed on Kuaishou, one of China's largest short-video platforms, serving hundreds of millions of users daily.
\end{abstract}

%
%



\received{20 February 2007}
\received[revised]{12 March 2009}
\received[accepted]{5 June 2009}

\maketitle

\section{Introduction}

\begin{figure*}[t]
\centering
\includegraphics[width=0.9\textwidth]{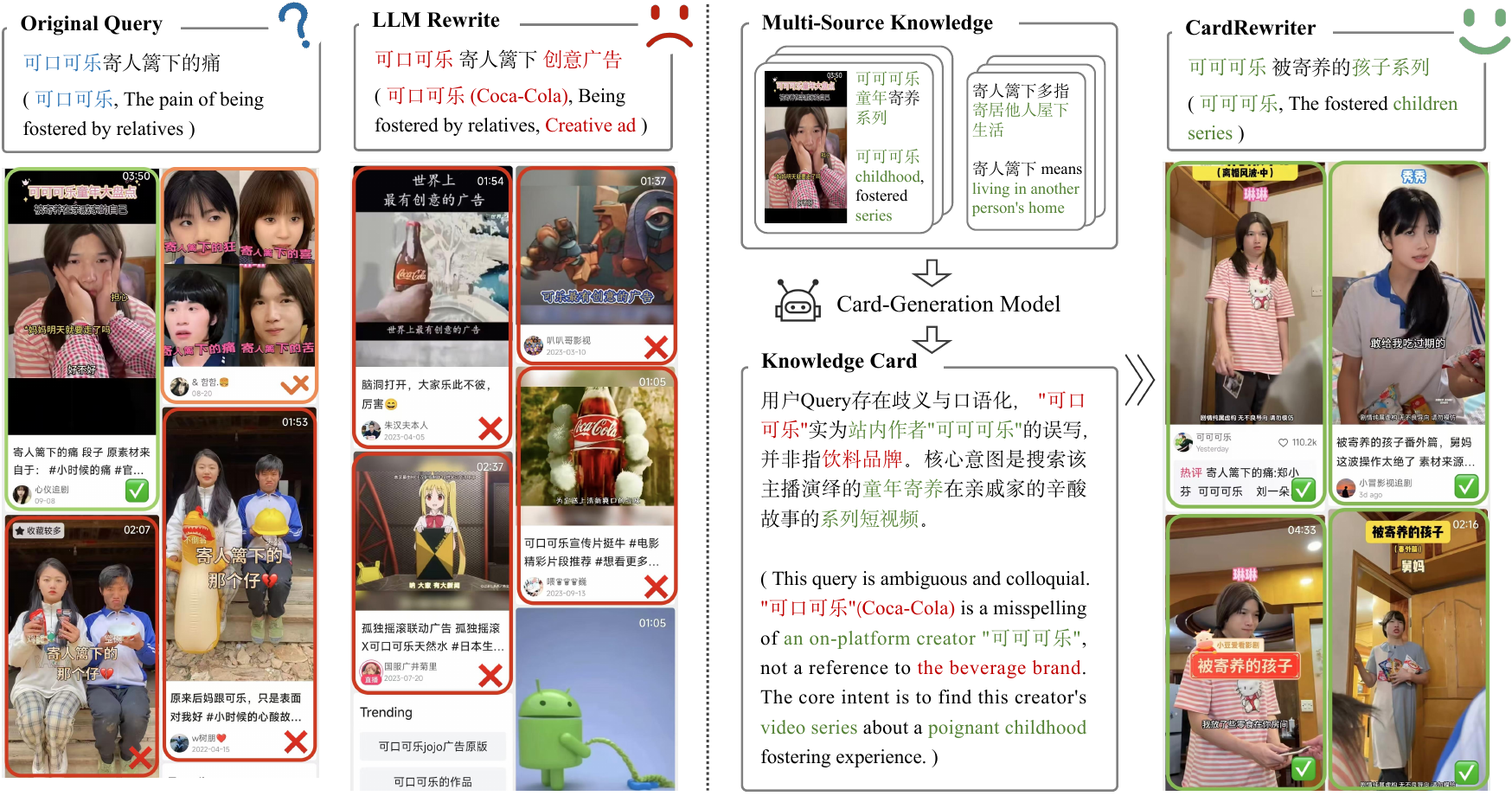}
\caption{
    An example of query rewriting on the short-video platform. (a) Original Query: Fails to retrieve relevant results; (b) LLM Rewrite: Misinterprets semantics and yields an ineffective rewrite; (c) CardRewriter: Leverages the generated knowledge card as search context to produce an accurate rewrite.
}
\label{intro_case}
\end{figure*}

In recent years, short-video platforms (e.g., Kuaishou, TikTok, Xiaohongshu) have become ubiquitous in daily life, enabling users to explore and interact with diverse content. 
Increasingly, these platforms also serve as search engines, allowing users to issue queries and receive ranked short-video results tailored to their information needs \cite{si2023search, sun2023kuaisar}. 
However, the diversity in users’ expressions and intents often gives rise to challenges such as spelling errors, missing terms,  and semantic ambiguity \cite{peng2024large, feng2025complicated}, especially for long-tail queries. 
Such queries frequently yield unsatisfactory search results, forcing users to repeatedly reformulate their requests or even abandon the search. 
Solving this problem is therefore critical to improving user satisfaction.

To address this challenge, query rewriting has become an intuitive and widely used solution \cite{liu2020query}, capable of correcting linguistic errors \cite{ye2023improving, kakkar2023search}, supplementing missing keywords \cite{naseri2021ceqe, wang2023query2doc, mackie2023generative}, and matching query style to available content \cite{dai2024enhancing, peng2024large, feng2025complicated}.
Existing methods fall into two main paradigms:
(I) \textit{Embedding-based methods} retrieve semantically similar queries to enrich the original, thereby increasing relevant content \cite{xiao2019weakly, li2022query, wang2023learning}. 
Despite this, they struggle with long-tail queries due to scarce semantic matches.
(II) \textit{Generative methods} directly rewrite long‑tail queries for greater semantic clarity and completeness \cite{zuo2022context, qiu2021query}.
Recently, large language model (LLM)-based rewriting has emerged as the dominant approach by using reinforcement learning to align user queries with platform-specific expression styles \cite{zuo2025value, peng2024large, dai2024enhancing, chen2025iterqr, feng2025complicated}.

While LLM-based query rewriting has shown strong performance in refining long-tail queries within e-commerce scenarios \cite{peng2024large, feng2025complicated}, extending these methods to short-video platforms remains challenging. 
Such platforms contain vast amounts of proprietary content, such as short videos, live streams, micro dramas, and user social networks, that LLMs have not been exposed to during pre-training, often leading to rewriting failures.
As illustrated in Figure~\ref{intro_case}, when a user attempts to search for the video series \textit{“The Fostered Children”} by the creator \textit{“Coco-Cola”} but mistakenly types \textit{“Coca-Cola”}, the LLM erroneously associates the query with the beverage brand and appends \textit{“creative ad”} to the original query. 
Consequently, the returned videos are entirely irrelevant to the user’s original search intent.

To address the above challenge, we propose CardRewriter, a retrieval-augmented framework that leverages platform-specific knowledge to enhance the quality of LLM-based query rewriting.
Given an input query, our approach first retrieves multi-source knowledge from short-video platforms, incorporating multi-modal content from top-$k$ videos retrieved through the original query and each query in the similar high-supply query set, respectively, along with relevant open-domain documents. 
This collected knowledge is then summarized into a concise knowledge card aligned with the input query, which is then integrated into the rewriting process to guide the LLM to better understand the search context and produce more effective query rewrites.
As shown in Figure \ref{intro_case}, the knowledge card accurately corrects the user's misspelling, thus retrieving more videos from the creator "Coco-Cola"'s series "The Fostered Children".
To ensure high-quality knowledge cards and rewrites, we adopt a two-phase training paradigm combining supervised fine-tuning (SFT) and group relative policy optimization (GRPO), guided by a custom reward system balancing both query relevance and retrieval effectiveness.

To evaluate the effectiveness of CardRewriter in rewriting long-tail queries on short-video platforms, we conduct both offline and online evaluations.
Offline results show that our approach improves the semantic relevance between rewritten and original queries and retrieves more videos matching users’ information needs.
In contrast, directly integrating multi-source knowledge introduces substantial noise and often distorts intent.
Our SFT+GRPO training strategy with a tailored reward system further boosts the quality of knowledge cards and rewritten queries. 
Online A/B tests confirm significant gains in long-view rate (LVR) and click-through rate (CTR), alongside reduced initiative query reformulation rate (IQRR) across both covered and full traffic.
Since its deployment on Kuaishou in September 2025, CardRewriter has greatly enhanced the search experience for hundreds of millions of users.

Our \textbf{main contributions} are as follows:
\begin{itemize}
    \item We propose CardRewriter, a framework that injects short-video platform knowledge into long-tail query rewriting, summarizing such knowledge into concise knowledge cards that guide LLM rewrites.
    \item We design a two-stage training framework that integrates SFT and GRPO, optimized by a custom reward system balancing relevance and effectiveness.
    \item Both offline and online experiments confirm that our method improves knowledge utilization and rewriting effectiveness, showing practical gains in real-world deployments.
\end{itemize}

\section{Related Work}

\subsection{Query Rewriting}
As an upstream task in information retrieval, query rewriting aims to refine the original query to better capture the user’s search intent \cite{liu2020query, song2024survey}.
Its objective is to improve the accuracy of search results and thereby enhance the overall search experience. 
Existing approaches can be broadly divided into embedding-based and generative methods.

\textbf{Embedding-based methods} 
treat query rewriting as a retrieval task by identifying similar queries from a pre-built set to augment the original query and improve recall.
For instance, \citet{li2022query} employs a contrastive learning architecture in a “retrieval–ranking–rer\\anking” pipeline to generate personalized rewrites.
\citet{xiao2019weakly} jointly trains query rewriting and semantic matching on weakly labeled data, enhancing both tasks through iterative co-training.
\citet{gamzu2020query} utilizes a search index to generate alternative queries, especially for voice search, and applies learning-to-rank to select the best rewrite.
Despite effectively boosting recall for common queries, these methods struggle with long-tail queries due to the scarcity of similar candidates.

\textbf{Generative methods} formulate query rewriting as a text generation task, directly producing revised queries without retrieval. \citet{qiu2021query} proposes a cycle-consistent training approach that jointly optimizes query-to-title and title-to-query models, while \citet{zuo2022context} constructs a session graph to capture historical interactions and integrate them via graph attention.
Recently, LLM-based methods have excelled at rewriting long-tail queries in e-commerce.
For example, \citet{peng2024large} designs a three-stage framework to align long-tail queries with product descriptions; 
\citet{dai2024enhancing} adapts LLMs to domain-specific patterns through pre-training, and \citet{zuo2025value} introduces a value-aware LLM that employs a weighted trie to generate high-value bidwords, improving both relevance and revenue.
Nevertheless, on short-video platforms, LLMs still face challenges in handling long-tail queries targeting proprietary content, often failing to fully capture user intent.

\subsection{Retrieval-augmented Generation}
Retrieval-augmented Generation (RAG) enhances response reliability by retrieving relevant information and leveraging it to generate factual, high-quality answers.\cite{gao2023retrieval, li2022survey, mei2025survey}.
\citet{guu2020retrieval} augments language models with a latent retriever over large corpora, while \citet{yan2024corrective} evaluates retrieval quality and reduces noise via a decompose–recompose algorithm.
\citet{gong2024cosearchagent} leverages LLMs to comprehend multi-party dialogues for web-augmented responses, and \citet{jiang2023active} proactively retrieves content by predicting future sentences.
Beyond individual components, \citet{gao2024smartrag} jointly tunes retrieval, rewriting, and generation with reinforcement learning, and \citet{chen2025improving} formulates the pipeline as a cooperative multi-agent task with shared rewards.
In this work, we collect multi-source knowledge for long-tail queries from short-video platforms and integrate it into query-relevant knowledge cards. 
These cards serve as enriched contexts to achieve better query rewriting.

\section{CardRewriter}
In this section, we detail how CardRewriter integrates knowledge from short-video platforms into query rewriting (Section \ref{rewriting_workflow}), present the training framework (Section \ref{training_framework}), and introduce the real-world deployment strategy (Section \ref{serving}).

\subsection{Rewriting Workflow}
\label{rewriting_workflow}
\begin{figure*}[t]
\centering
\includegraphics[width=\textwidth]{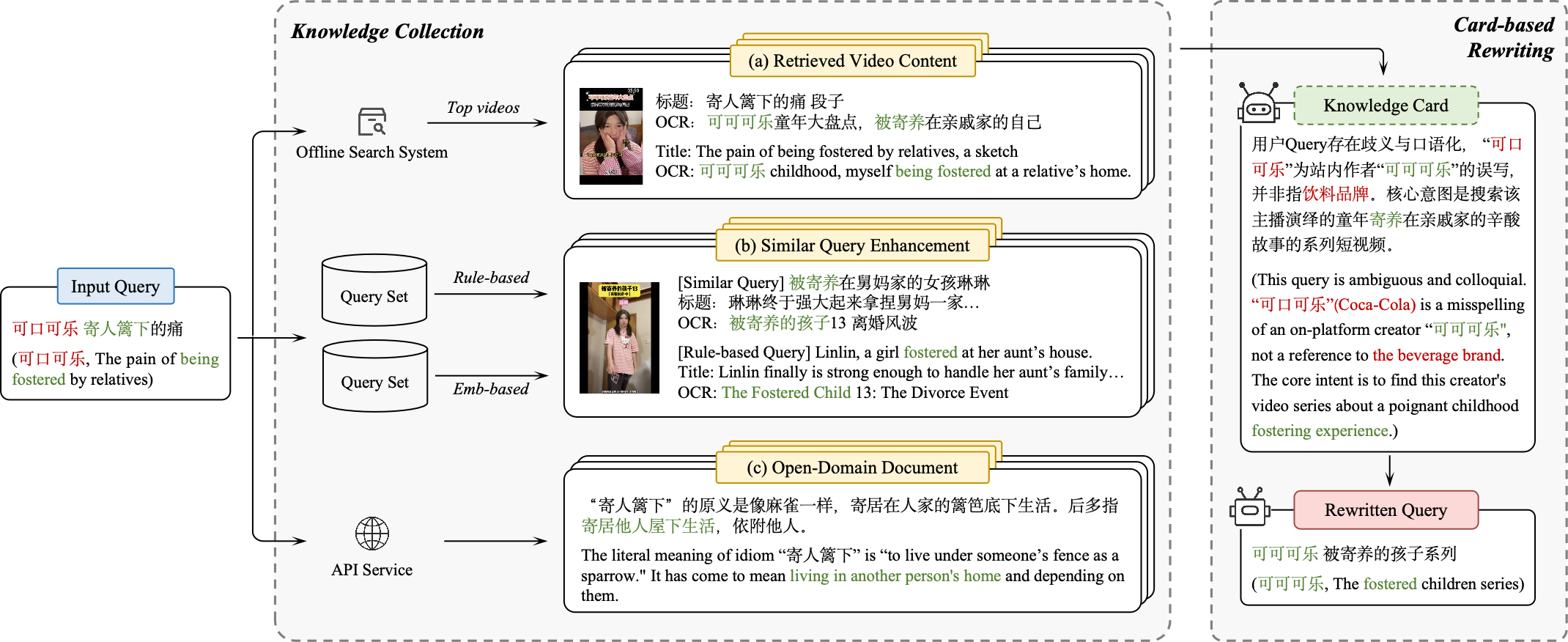}
\caption{
The overall workflow of CardRewriter. Given an input query, CardRewriter retrieves multi-source knowledge from short-video platforms, summarizes it into a concise knowledge card relevant to the query, and then leverages this card to better interpret user intent and refine the query.
}
\label{cardrewriter}
\end{figure*}

To leverage short-video platform knowledge for query rewriting, as shown in Figure \ref{cardrewriter}, CardRewriter operates in two stages:
(I) Knowledge Collection: aggregating the top-$k$ videos retrieved for each long-tail query and its corresponding similar high-supply queries, and further enriching them with relevant open-domain documents;
(II) Card-Based Rewriting: summarizing the collected knowledge into knowledge cards that help the LLM to generate effective rewrites.
Detailed prompts used for CardRewriter are provided in Appendix \ref{prompts}.

\subsubsection{Knowledge Collection}
\label{collection}
LLMs often struggle to understand queries on short-video platforms when the search intent points to proprietary content, resulting in ineffective query rewriting. 
To address this, we collect platform-specific knowledge to enhance LLMs’ understanding of such queries.
For the given query $x$, we feed it into the Kuaishou search system to retrieve the top-$k$ relevant videos $V = \{v_1, v_2, \ldots, v_k\}$, and extract multi-modal knowledge $M_{\text{in}} = \{m_1, m_2, \ldots, m_k\}$ from them.
Specifically, for each video $v_i$, we extract content potentially useful for understanding $x$, including:
(I) Visual content $v_i^{\text{vision}}$: three key frames of the video;
(II) Textual content $v_i^{\text{text}}$: the title, caption, text on the cover (OCR), author name, and background music (BGM) of the video.
Formally, this can be represented as:
\begin{equation}
    v_i^{\text{vision}} = \{v_i^{\text{key}_1}, v_i^{\text{key}_2}, v_i^{\text{key}_3}\} 
\end{equation}
\begin{equation}
    v_i^{\text{text}} = \{v_i^{\text{title}}, v_i^{\text{caption}}, v_i^{\text{ocr}}, v_i^{\text{author}}, v_i^{\text{bgm}}\} 
\end{equation}
\begin{equation}
    m_i = \{v_i^{\text{vision}}, v_i^{\text{text}}\}
\end{equation}
Where $m_i$ denotes the multi-modal knowledge extracted from video $v_i$, consisting of visual content $v_i^{\text{vision}}$ and textual content $v_i^{\text{text}}$.

Moreover, to address the low relevance of top-$k$ retrievals for certain long-tail queries, we first construct two high-quality query sets $Q_{\text{good}}^{(1)}$ and $Q_{\text{good}}^{(2)}$ from Kuaishou search logs.
We then retrieve the top-$l$ queries similar to the original query $x$ using two methods separately:
(I) Rule-based matching ($\text{Q2Q}$): select queries from $Q_{\text{good}}^{(1)}$ that share lexical overlap with $x$ and have intersecting retrieved video lists, then rank them by the number of identical videos.
(II) Embedding-based matching ($\text{EMB}$): compute embedding similarity between $x$ and each query in $Q_{\text{good}}^{(2)}$, then rank them by similarity.
The set of similar queries can be formalized as:
\begin{equation}
Q_{\text{sim}} = \text{Set}(\text{Q2Q}(x, Q_{\text{good}}^{(1)}) \cup \text{EMB}(x, Q_{\text{good}}^{(2)}))
\end{equation}
Here, $Q_{\text{sim}}$ denotes the deduplicated query set similar to $x$.

For each query $q \in Q_{\text{sim}}$, we extract its multi-modal knowledge $M_q$ from the top-$k$ retrieved videos in the same manner as $M_{\text{in}}$. The enriched knowledge set $M_{\text{sim}}$ is then defined as:
\begin{equation}
M_{\text{sim}} = \bigcup_{q \in Q_{\text{sim}}} M_q
\end{equation}

Finally, we obtain open-domain documents $M_{\text{ex}}$ through the self-built API service, serving as an additional knowledge source when relevant videos are scarce. 
We aggregate all collected knowledge and remove duplicates, yielding:
\begin{equation}
M = \{\text{Set}(M_{\text{in}} \cup M_{\text{sim}}), M_{\text{ex}}\}
\end{equation}
where $M$ represents the multi-source knowledge collected for query $x$ to support query rewriting on short-video platforms.

\subsubsection{Card-based Rewriting}
The multi-source knowledge collected in Section \ref{collection} often suffers from structural inconsistencies, excessive noise, and limited relevance to the input queries \cite{fang2024enhancing, cuconasu2024power}.
To mitigate these issues, we investigate how to effectively integrate, denoise, and exploit the collected knowledge.
We first employ a card generation model $\mathcal{C}_\theta$ to summary such knowledge into a concise and informative knowledge card that highlights content directly related to user queries.
Subsequently, a query rewriting model $\mathcal{G}_\theta$ leverages the generated card to rewrite the input query, thereby enhancing its understanding of the search context.
This design enables accurate rewriting without resorting to resource-intensive pre-training \cite{dai2024enhancing}, which is costly and challenging to update frequently.
Both $\mathcal{C}_\theta$ and $\mathcal{G}_\theta$ utilize LLM as their foundation, possessing strong capabilities in contextual understanding and text generation. 
The overall process can be expressed as:
\begin{align}
    y = \mathcal{G}_\theta(x, c), \quad c = \mathcal{C}_\theta(x, M)
\end{align}
Where $y$ denotes the rewritten query and $c$ presents the generated knowledge card.

\subsection{Training Framework}
\label{training_framework}
\begin{figure*}[t]
\centering
\includegraphics[width=\textwidth]{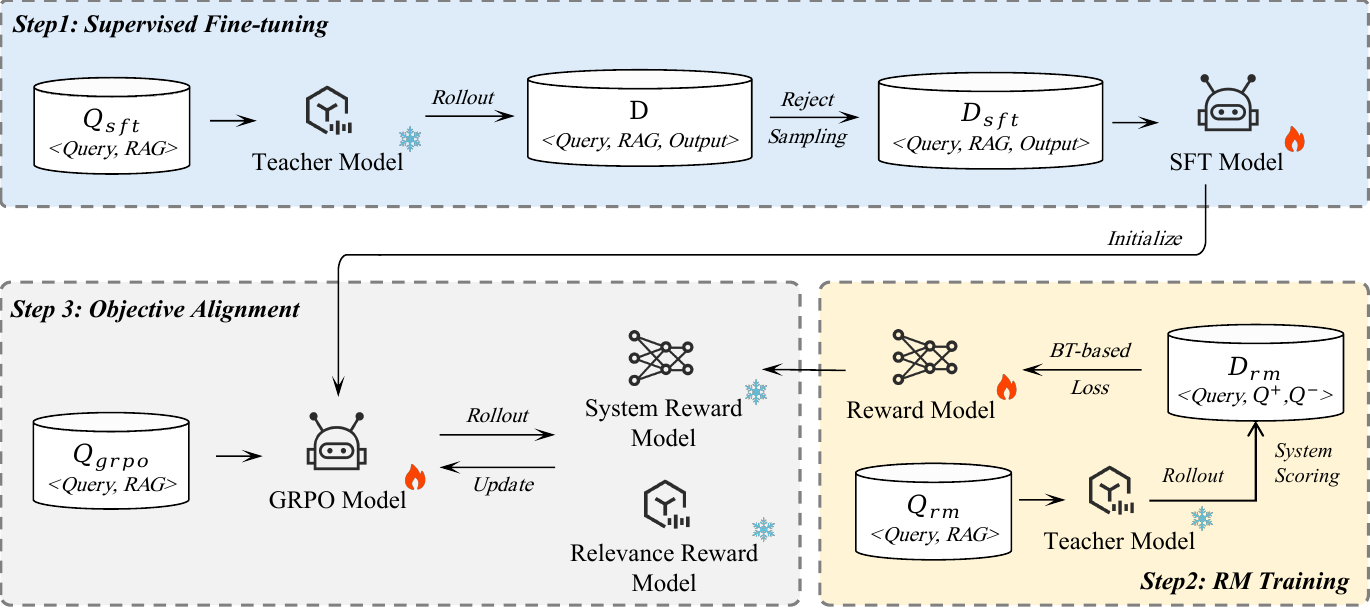}
\caption{Training Strategy.}
\label{training}
\end{figure*}

To optimize both the card generation and query rewriting models, as illustrated in Figure \ref{training}, we employ a two-stage training pipeline for each, consisting of supervised fine-tuning (SFT) followed by preference alignment through group relative policy optimization (GRPO). 
Furthermore, we design a reward system that balances the semantic relevance between the original query and the model’s output, as well as the latter’s influence on the quality of the retrieved video list, thereby guiding the GRPO training process.

\subsubsection{Task Formulation}
Both tasks can be unified as $y \leftarrow (x, K)$, where $x$ is the input query.
For card generation, $K$ denotes the multi-source knowledge collected for $x$, and $y$ is the resulting knowledge card.
For query rewriting, $K$ corresponds to the generated knowledge card, and $y$ is the rewritten query.
We further introduce a rewritten query variable, $rq$, which is used for three key purposes: (I) data filtering during the SFT stage, (II) training the reward model that predicts system preference, and (III) reward computation in the GRPO stage.
In the card generation task, $rq$ is obtained by first generating a knowledge card and then perform rewriting, whereas in the query rewriting task, it is generated directly.
We train the card generation model $\mathcal{C}_\theta$ and the query rewriting model $\mathcal{G}_\theta$ through the SFT+GRPO pipeline, with both two models represented as $\pi_{SFT}$ and $\pi_{GRPO}$ in these two stages respectively.

\subsubsection{Supervised Fine-tuning}
\label{cg_sft}
High-quality training data is essential for the SFT stage of both tasks.  
To this end, we construct a large query set $Q_{sft}$ from Kuaishou’s search logs and collect the corresponding knowledge $K$ for each query $x \in Q_{sft}$, which presents either multi-source knowledge or a knowledge card.  
For each query, we generate an output $y$ conditioned on its knowledge $K$, forming the initial dataset:  
$$D = \{(x, K, y) \mid x \in Q_{sft}\}.$$  

To further improve data quality, we filter $D$ according to two criteria:  
(I) Semantic Relevance, assessed by a judge model $\mathcal{R}_{\text{Rel}}$\footnote{\url{https://huggingface.co/Qwen/Qwen3-235B-A22B}}, which evaluates query–card relevance and query–rewrite relevance in the two tasks, respectively; and  
(II) System Preference, denoted as $\text{SYS}(\cdot)$, which first compares the hitrate score between the rewritten query $rq$ and the original query $x$, and if equal, compares their increment score.  
The computation of these metrics is detailed in Section~\ref{evaluation}.  
The resulting SFT dataset is defined as:
\begin{equation} \begin{split} D_{sft} &= \{(x, K, y) \mid (x, K, y) \in D, \\ &\quad \mathrm{SYS}(x, rq) = 1,\ \mathcal{R}_{\text{Rel}}(x, y) = 1 \}, \end{split} \end{equation}
where $\mathrm{SYS}(x, rq)=1$ denotes the rewritten query is better than the original query from the search system view, $\mathcal{R}_{\text{Rel}}(x, y) = 1$ indicates semantic alignment.

We fine-tune the SFT model $\pi_{SFT}$ as a conditional sequence generator using token-level cross-entropy:
\begin{equation}
    \mathcal{L}_{\text{SFT}}(\theta) = -\mathbb{E}_{(x, K, y) \sim D_{sft}} \left[ \log \pi_{SFT}(y \mid x, K) \right].
\end{equation}

\subsubsection{Reward System}
\label{rm}
We present the construction of our self-designed reward system, denoted as $\mathcal{R}_{\text{Overall}}$, which aims to balance query relevance and retrieval effectiveness. 
Specifically, $\mathcal{R}_{\text{Overall}}$ consists of two components: $\mathcal{R}_{\text{Rel}}$, which measures semantic relevance, and $\mathcal{R}_{\text{Sys}}$, which simulates system-level preferences.
Since direct system preference scores are not readily accessible during model training, we build Bradley–Terry (BT)-based reward model $\mathcal{R}_{\text{Sys}}$ \cite{sun2024rethinking} to approximate system evaluations and provide feedback for subsequent GRPO optimization. 
To construct the training data for this reward model, we first sample a large query set $Q_{rm}$ from Kuaishou’s search logs. 
For each query $x \in Q_{rm}$, we apply the rewriting pipeline described in Section \ref{rewriting_workflow} to generate two candidate rewrites. These rewrites are then evaluated by the offline search system through $\text{SYS}(\cdot)$, and construct the reward model dataset $D_{rm} = \{(x, rq^+, rq^-)\mid x \in Q_{rm}\}$, where $rq^+$ and $rq^-$ denote the preferred and less-preferred rewrites, respectively. Following the BT formulation, the preference probability is:
\begin{equation}
P_\theta(rq^+ \succ rq^- \mid x) =
\frac{\exp(\mathcal{R}_{\text{Sys}}(x,rq^+))}
{\exp(\mathcal{R}_{\text{Sys}}(x,rq^+)) + \exp(\mathcal{R}_{\text{Sys}}(x,rq^-))}.
\end{equation}

The $\mathcal{R}_{\text{Sys}}$ is trained by minimizing the negative log-likelihood of
observed preferences:
\begin{equation}
\mathcal{L}_{RM}(\theta) = - \mathbb{E}_{(x,rq^+,rq^-) \sim D_{rm}}
\Big[ \log P_\theta(rq^+ \succ rq^- \mid x) \Big].
\end{equation}

Finally, we combine $\mathcal{R}_{\text{Sys}}$ and $\mathcal{R}_{\text{Rel}}$ to form $\mathcal{R}_{\text{Overall}}$:
\begin{equation}
\mathcal{R}_{\text{Overall}} = 
\begin{cases}
\mathcal{R}_{\text{Sys}}, & \text{if } \mathcal{R}_{\text{Sys}} > 0 \\
0.1, & \text{if } \mathcal{R}_{\text{Sys}} = 0 \text{ and } \mathcal{R}_{\text{Rel}} > 0 \\
0, & \text{if } \mathcal{R}_{\text{Sys}} = \mathcal{R}_{\text{Rel}} = 0
\end{cases}
\end{equation}

\subsubsection{Objective alignment}
Equipped with the SFT model $\pi_{SFT}$ and the overall reward function $\mathcal{R}_{\text{Overall}}$, which jointly balance relevance and system preference, we employ the Group Relative Policy Optimization (GRPO) algorithm~\cite{shao2024deepseekmath} to align long-tail queries with the descriptive style of proprietary short-video content. 
Specifically, we construct the training dataset $D_{grpo}$ by collecting real user queries from Kuaishou’s search logs. 
For each query $x \in D_{grpo}$, a group of $G$ rollout trajectories $y = \{y_i\}_{i=1}^G$ is generated using the 
previous policy $\pi_{old}$. The current policy model $\pi_{GRPO}$ is optimized by maximizing the following objective:
\begin{equation}
\begin{aligned}
\mathcal{J}_{GRPO}(\theta) &= \mathbb{E}_{\substack{(x,K) \sim D_{grpo}, \; \{y_i\}_{i=1}^G \\ y_i \sim \pi_{\mathrm{old}}(\cdot \mid x, K)}}\Bigg[ \frac{1}{G} \sum_{i=1}^G \min \Big( \frac{\pi_{GRPO}(y^i | x, K)}{\pi_{\theta_{old}}(y^i | x, K)} \hat{A}_i, \\
&\quad \text{clip} \big( \frac{\pi_{GRPO}(y^i | x, K)}{\pi_{\theta_{old}}(y^i | x, K)}, 1 - \epsilon, 1 + \epsilon \big) \hat{A}_i \Big) - \beta \text{KL} \big[ \pi_{GRPO} \big\| \pi_{ref} \big] \Bigg]
\end{aligned}
\end{equation}
where $\epsilon$ denotes the clipping ratio, and $\hat{A}_i$ represents the advantage of the $i$-th rewritten query $rq_i$, computed based on the overall reward $\mathcal{R}_{\text{Overall}}$ across all rewrites within the same group.

\begin{figure}[t]
\centering
\includegraphics[width=\columnwidth]{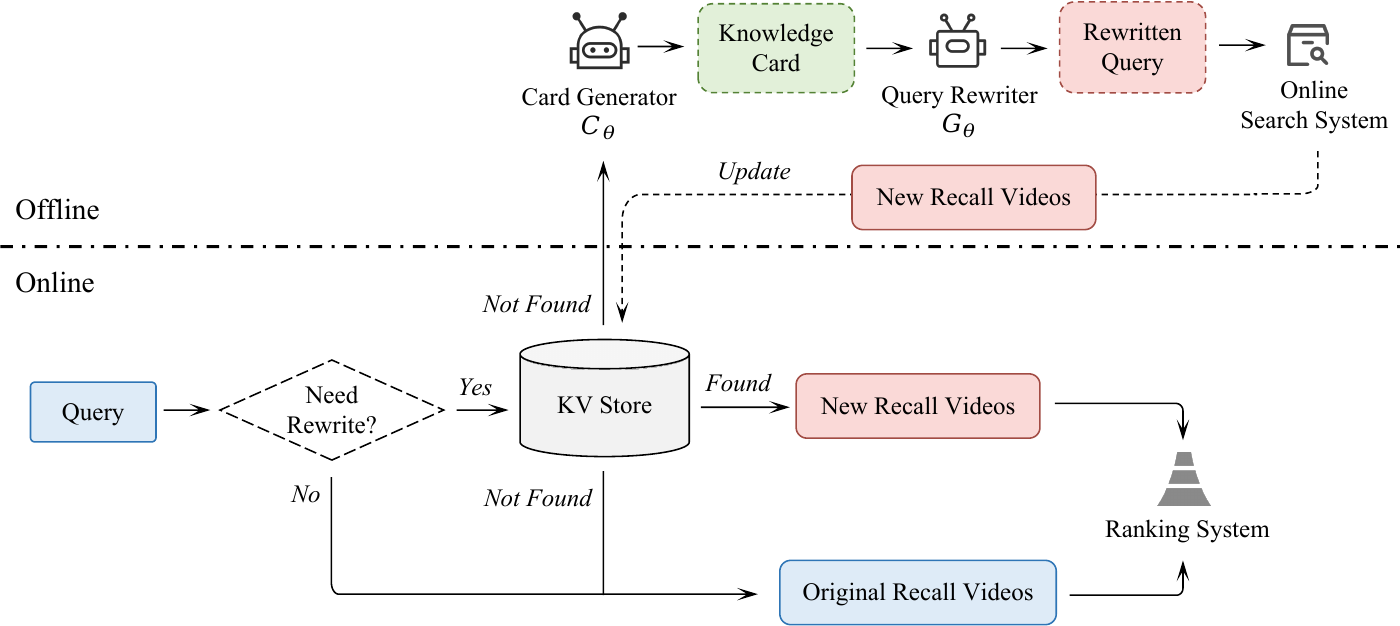}
\caption{Development Strategy.}
\label{develop}
\vspace{-1em}
\end{figure}

\begin{table*}[t]
\centering
\caption{Overall performance of CardRewriter with multiple baselines. CardRewriter demonstrates the strongest performance across most metrics. We highlight the highest score in bold and the second-highest score with underlines.}
\begin{tabular}{l l c c c c}
\toprule
\textbf{Method} & \textbf{Knowledge} & \textbf{QR-Rel} & \textbf{Increment} & \textbf{Hitrate@50} & \textbf{Hitrate@300} \\
\midrule
Original Query & - & - & - & 31.40\% & 53.07\% \\
Previous \cite{peng2024large, dai2024enhancing, feng2025complicated} & - & 78.42\% & 66.82\% & 36.56\% & 59.96\% \\
\midrule
\multirow{3}{*}{Prompt} 
& - & 69.72\% & 30.17\% & 29.04\% & 48.26\% \\
& Naive RAG & 62.78\% & 42.29\% & 29.49\% & 48.53\% \\
& Card RAG & 74.18\% & 48.71\%  & 32.72\% & 54.76\% \\
\midrule
\multirow{3}{*}{SFT} 
& - & 76.35\% & 58.08\% & 33.86\% & 58.15\% \\
& Naive RAG & 73.68\% & 64.69\% & 33.95\% & 59.18\% \\
& Card RAG & 84.67\% & 66.88\% & 39.57\% & 66.18\% \\
\midrule
\multirow{3}{*}{SFT+DPO} 
& - & 77.36\% & 65.42\% & 37.96\% & 62.39\% \\
& Naive RAG & 74.93\% & 67.17\% & 38.72\% & 63.03\% \\
& Card RAG & \textbf{86.23\%} & 70.53\% & \underline{42.18\%} & \underline{70.32\%} \\
\midrule
\multirow{3}{*}{SFT+GRPO} 
& - & 78.98\% & 65.19\% & 41.68\% & 65.71\% \\
& Naive RAG & 74.28\% & \underline{70.86\%} & 41.05\% & 65.63\% \\
& Card RAG (CardRewriter) & \underline{85.73\%} & \textbf{74.17\%} & \textbf{46.64\%} & \textbf{76.04\%} \\
\bottomrule
\end{tabular}
\label{tab:results}
\end{table*}

\subsection{Online Serving}
\label{serving}
Due to the large parameter size and the auto-regressive nature of LLMs, directly deploying CardRewriter under Kuaishou’s stringent low-latency search requirements is infeasible.
To address this limitation, we adopt a near-line deployment strategy \cite{peng2024large, dai2024enhancing}, in which query rewriting is performed offline for a targeted subset of queries.
The selection criteria are: (i) Average daily searches between 5 and 300 in the past 7 days; (ii) Query intent not limited to a username; and (iii) poor retrieval performance, characterized by low average relevance, low click-through rates, and high query reformulation rates.
This selection covers approximately 15–20\% of daily search traffic, corresponding to around 5 million queries.
For each selected query, we first collect multi-source knowledge and summarize it into a knowledge card via $\mathcal{C}_\theta$. 
The query is then rewritten by $\mathcal{G}_\theta$ based on the generated knowledge card, and the retrieved most relevant videos are cached in an online key–value (KV) store.
Storing videos instead of rewriting queries can eliminate the impact of personalization while saving online inference resources.
During the online stage, as illustrated in Figure~\ref{develop}, each incoming query first checks the KV store: if a match is found, the cached videos are appended to the original video list; otherwise, the offline rewriting process is repeated. 
Cached entries that result in misses expire after seven days, or sooner if their relevance or click-through rate exceeds predefined thresholds.

\section{Experiments}

\subsection{Datasets}
\subsubsection{Training Dataset}
We construct three datasets for card generation, query rewriting, and reward modeling.
(I) Card generation: We sample 200k queries from Kuaishou’s search logs, retrieve multi-source knowledge for each, and generate 8 cards per query, resulting in 1.6M \textit{<query, knowledge, card, rewrite>} quadruples. Through rejection sampling, we retain around 30k high-quality \textit{<query, knowledge, card>} triples for SFT. Additionally, we collect 60k queries for GRPO training.
(II) Card-based rewriting:  We collect 400k queries and follow the same pipeline, producing 3.2M \textit{<query, card, rewrite>} triples. After filtering, 50k high-quality pairs remain for SFT training. We further sample 100k queries for GRPO training.
(III) Reward modeling: We gather 150k queries, generate multiple rewrites for each, and assign system preference scores (Section~\ref{rm}) derived from the Kuaishou search system. This process yields approximately 240k \textit{<query, good\_rewrite, bad\_rewrite>} tuples, which are used to train a reward model that predicts system preferences.

\subsubsection{Test Dataset}
To evaluate model performance across different tasks, we construct task-specific test sets: 10k queries for card generation, 10k queries for reward modeling, and 15k queries for query rewriting, all sampled from Kuaishou search logs.

\subsection{Evaluation}
\label{evaluation}
\subsubsection{Offline Metrics}
We employ Rel, Increment, and Hitrate@$K$ as offline evaluation metrics, each capturing a distinct dimension of model performance:
(I) Rel (Relevance): Evaluates the semantic quality of model outputs.
For card generation, it measures how relevant the generated knowledge card is to the original query, referred to as QC-Rel.
For query rewriting, it assesses i) the relevance between the rewritten and original queries, and ii) whether the rewriting effectively integrates the knowledge card, denoted as QR-Rel.
(II) Increment (Retrieval Expansion): Measures the model’s capability to expand retrieval coverage by quantifying the relative improvement in recall when using both the original and rewritten queries, compared to using the original query alone.
(III) Hitrate@$K$ (User Satisfaction): Reflects the proportion of cases in which the rewritten query successfully retrieves at least one video within the top-$K$ results that aligns with the user’s intent.
More details about the above metrics can be found in Appendix \ref{appendix_metrics}.

\subsubsection{Online Metrics}
We introduce three core online metrics, LVR, IQRR, and CTR, which best reflect the user's search experience on short-video platforms and are used to evaluate the online performance of our method.
These metrics are defined as follows: \textit{LVR = long-view rates, IQRR = initiative query reformulation rates  and CTR = click-through rates.}

\subsection{Offline Experiments}
\subsubsection{Main Results}
In our experiments, we compare CardRewriter with prompt-based, SFT-based, and SFT+DPO-based rewriting methods, as well as knowledge-enhanced baselines that either exclude augmentation or directly integrate multi-source knowledge (Naive RAG). 
We further adapt the core modules of CSA-QR~\cite{feng2025complicated} to the Kuaishou platform for comparison. 
Implementation details of CardR\\ewriter are provided in Appendix \ref{details}. 
Key observations based on Table~\ref{tab:results} are summarized as follows:

\textbf{Directly injecting multi-source knowledge offers limited benefits for query rewriting.}
We evaluate a rewriting approach that integrates multi-source knowledge against a baseline without enhancement.
Results show that directly injecting such knowledge reduces semantic alignment between rewritten and original queries across all three baseline methods and our SFT+GRPO model.
The degradation mainly arises from retrieval-augmented rewriting, which lengthens queries and introduces noise that drifts from the original intent. Interestingly, naive knowledge injection raises the increment metric by retrieving additional unseen videos, but hitrate scores remain unstable, suggesting that few of these new results match the ground truth.

\begin{table}[t]
\centering
\caption{Ablation study on CardRewriter with different types of knowledge sources. Both knowledge cards and rewritten queries are derived through a prompt-based method.}
\resizebox{\columnwidth}{!}{
\begin{tabular}{l c c c}
\toprule
\textbf{Method} & \textbf{QC-Rel} & \textbf{Increment} & \textbf{Hitrate@300} \\
\midrule
CardRewriter & 91.16\% & 46.24\% & 51.36\% \\
\midrule
\multicolumn{4}{l}{\textit{Modal Type}} \\
\quad w/o vision & 89.37\% & 45.01\% & 50.14\% \\
\quad w/o textual & 86.18\% & 42.38\% & 45.62\% \\
\midrule
\multicolumn{4}{l}{\textit{Video Retrieval}} \\
\quad w/o rel-videos & 78.27\% & 40.42\% & 47.16\% \\
\quad w/o rule sim & 93.24\% & 44.12\% & 50.37\% \\
\quad w/o emb sim & 92.86\% & 44.89\% & 50.86\% \\
\midrule
\multicolumn{4}{l}{\textit{Open Domain}} \\
\quad w/o ext & 92.06\% & 45.85\% & 50.31\% \\
\bottomrule
\end{tabular}}
\label{tab:ablation}
\end{table}

\textbf{Summarizing multi-source knowledge into knowledge cards significantly improves rewriting performance.}
Compared to directly injecting multi-source knowledge, which often introduces irrelevant or even conflicting content, summarizing it into knowledge cards delivers significant improvements.
For instance, SFT+DPO with Card RAG reaches a QR-Rel of 86.23\% and SFT+GRPO with Card RAG (CardRewriter) achieves a Hitrate@300 of 76.87\%, both significantly outperforming their Naive RAG counterparts.
Across all rewriting methods, the incorporation of knowledge cards not only markedly enhances the semantic relevance between rewritten and original queries, but also expands the scope of retrieved results and increases the likelihood of covering ground-truth videos.
The results demonstrate that knowledge cards effectively mitigate the shortcomings of Naive RAG and serve as a powerful medium for grounding LLM rewrites in platform-specific knowledge, enabling more precise retrieval.

\textbf{Different rewriting methods exhibit varying abilities to absorb external knowledge.}
Injecting Knowledge cards consistently surpass naive knowledge enhancement across all rewriting settings, though the degree of improvement depends on the rewriting strategy.
Prompt-based rewriting yields only marginal gains because it lacks targeted learning to integrate external knowledge effectively.
In contrast, SFT and SFT + DPO substantially enhance retrieval quality by aligning rewritten queries with contextual cues through supervised optimization.
Most notably, SFT + GRPO achieves the greatest overall improvement: its reward-guided training explicitly drives the model to exploit informative signals from knowledge cards.
Consequently, SFT + GRPO not only preserves semantic fidelity to the original query but also maximizes the utility of knowledge cards, leading to the strongest performance across QR-Rel and Hitrate metrics.

\subsubsection{Ablation Study on Multi-Source Knowledge}
To evaluate the contribution of different knowledge sources in CardRewriter, we perform an ablation study by removing each type of injected knowledge and examining the resulting performance. 
As shown in Table \ref{tab:ablation}, removing either visual or textual modalities from video content degrades rewriting quality, with textual information playing a more critical role. 
This is because missing or ambiguous content in long-tail queries can often be recovered from video titles and captions. 
From a retrieval perspective, rewriting benefits from both videos retrieved by the original query and those obtained through similar queries. 
Notably, even a few relevant videos retrieved by the form method are highly valuable, as they provide crucial evidence for building reliable knowledge cards and generating high-quality rewrites.
Finally, incorporating open-domain documents yields further gains by providing complementary information that enriches overall knowledge coverage.

\subsubsection{Results of Different Rewards}
\begin{figure}[t]
\centering
\includegraphics[width=0.9\columnwidth]{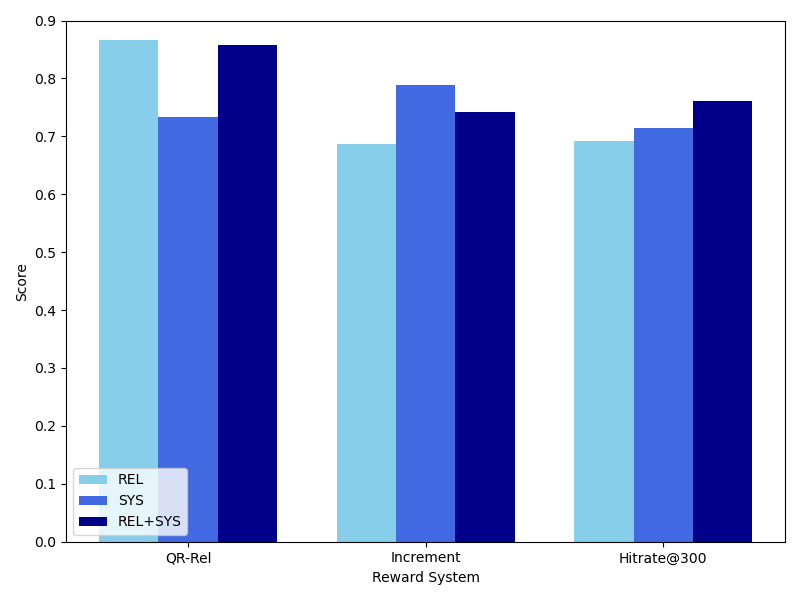}
\vspace{-1.5em}
\caption{Effectiveness of different rewards.}
\label{reward_fig}
\vspace{-1.5em}
\end{figure}

To assess how reward design shapes objective alignment, we vary the reward components in GRPO for the query rewriting task and analyze their impact on rewriting performance. 
Figure \ref{reward_fig} shows that using only the relevance reward (REL) achieves the highest query-rewrite relevance scores but fails to expand coverage to new videos that fully meet user needs. 
In contrast, relying solely on the system reward (SYS) retrieves many previously inaccessible videos yet often drifts from the original intent.
Combining the relevance and system rewards ensures high retrieval relevance while maximizing the number of user-prefer videos.
This demonstrates that our approach effectively balances query relevance and retrieval effectiveness, preserving user intent while enhancing overall search experience.

\begin{figure}[t]
\centering
\includegraphics[width=0.9\columnwidth]{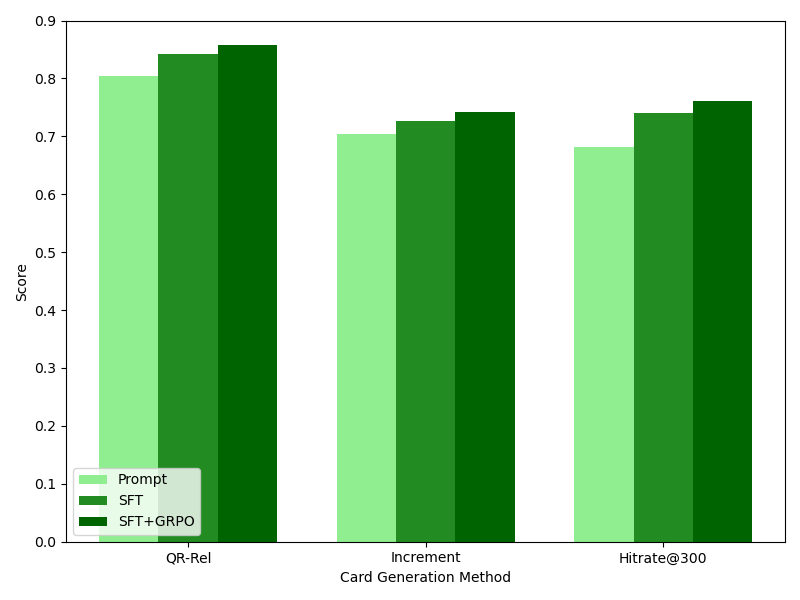}
\vspace{-1.5em}
\caption{Effectiveness of different card generation methods.}
\label{card_gen_fig}
\vspace{-1em}
\end{figure}

\begin{figure*}[t]
\centering
\includegraphics[width=0.95\textwidth]{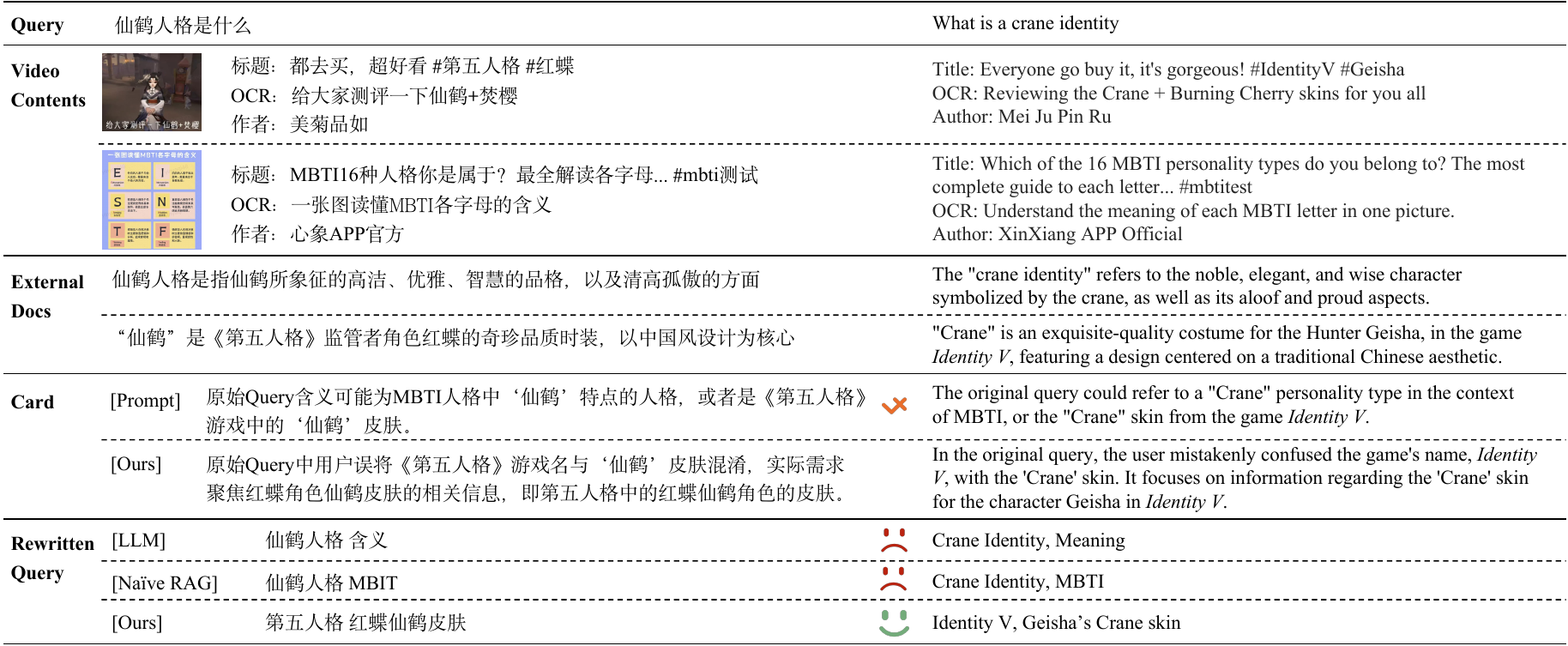}
\vspace{-1em}
\caption{
    Case Study of long-tail query on the short-video platform. CardRewriter remains reliable, producing accurate rewriting even when the collected knowledge is inconsistent.
}
\label{case}
\vspace{-1em}
\end{figure*}

\subsubsection{Results of Different Card Generation Methods}
To examine how card quality affects query rewriting, we evaluate the influence of knowledge cards generated by different methods. 
Specifically, we compare three approaches: prompt-based, SFT-based, and SFT+GRPO-based card generation. 
As shown in Figure \ref{card_gen_fig}, the prompt-based approach consistently performs worst across all metrics, underscoring the limits of relying solely on prompts. 
The SFT-based method substantially improves performance, demonstrating the benefit of training a dedicated card generation model over directly using prompts.
Building on this, the SFT+GRPO approach further refines card content via learned reward signals, achieving the best rewriting performance.
These results indicate that enhancing card quality through advanced training directly leads to higher-quality rewritten queries.

\subsubsection{Case Study}
As shown in Figure \ref{case}, our approach effectively rewrites queries aimed at retrieving proprietary content. 
In this example, direct rewriting by an LLM only paraphrases the query without resolving the ambiguous reference to "crane personality". 
Likewise, relying solely on multi-source knowledge fails due to conflicting evidence across videos and documents, for instance, whether the term denotes a crane-like personality or the "Crane Skin" in Identity V. 
In contrast, CardRewriter constructs a knowledge card that summarizes these sources, correctly interpreting the query as referring to the Identity V "Crane Skin". 
Moreover, prompt-based card generation alone cannot fully resolve such conflicts. 
By adding SFT + GRPO training, which explicitly optimizes retrieval for user-preferred videos, our method produces knowledge cards that align more closely with users’ actual search intent.

\begin{table}[t]
\centering
\caption{Online A/B test of CardRewriter on Kuaishou Search.}
\resizebox{0.8\columnwidth}{!}{
\begin{tabular}{l c c c}
\toprule
& \textbf{LVR ↑} & \textbf{IQRR ↓} & \textbf{CTR ↑}  \\
\midrule
Covered Traffic & +1.853\% & -2.630\% & +3.729\%  \\
Full Traffic & +0.235\% & -0.229\% & +0.342\%  \\
\bottomrule
\end{tabular}
}
\label{abtest}
\vspace{-1.5em}
\end{table}

\subsection{Online Experiments}
To assess the real-world online performance of CardRewriter, we conducted a 10-day deployment on Kuaishou Search.
Following the serving strategy described in Section \ref{serving}, the system is exposed to approximately 16\% of real traffic.
During this experiment, we monitor three key search metrics: LVR, IQRR, and CTR, which jointly capture user engagement and the overall quality of the search experience.
As summarized in Table \ref{abtest}, our approach delivers notable improvements on both the covered traffic and the full platform traffic.
On the covered traffic, LVR and CTR increase by 1.853\% and 3.729\% respectively, indicating a substantial improvement in user engagement and the relevance of search results. 
Meanwhile, IQRR decreases by 2.630\%, reflecting a marked reduction in users actively reformulating their queries.
Consistent improvements are also observed on full traffic, where LVR and CTR continue to rise and IQRR decline, demonstrating that CardRewriter not only performs well on a traffic subset but also robustly enhances the overall search experience.

\section{Conclusion}
In this work, we introduce CardRewriter, an LLM-based framework that leverages domain knowledge from short-video platforms to improve the rewriting of long-tail queries. 
For each query, we collect multi-source knowledge by retrieving top-$k$ videos from Kuaishou’s search system, aggregating relevant videos linked to similar high-supply queries from search logs, and supplementing with open-domain documents. 
This information is then integrated, denoised, and summarized into a concise knowledge card, which provides contextual guidance for accurate query understanding and high-quality rewriting.
To optimize CardRewriter, we employ a two-stage training strategy: supervised fine-tuning followed by group relative policy optimization, with a customized reward system balancing query relevance and retrieval effectiveness. 
Offline experiments demonstrate the effectiveness of our approach in rewriting queries aiming to retrieve proprietary content, while online A/B tests show significant gains in long-view rate (LVR) and click-through rate (CTR), as well as reductions in initiative query reformulation rate (IQRR), leading to overall better search performance.



\bibliographystyle{ACM-Reference-Format}
\bibliography{sample-base}


\begin{thebibliography}{33}


\ifx \showCODEN    \undefined \def \showCODEN     #1{\unskip}     \fi
\ifx \showISBNx    \undefined \def \showISBNx     #1{\unskip}     \fi
\ifx \showISBNxiii \undefined \def \showISBNxiii  #1{\unskip}     \fi
\ifx \showISSN     \undefined \def \showISSN      #1{\unskip}     \fi
\ifx \showLCCN     \undefined \def \showLCCN      #1{\unskip}     \fi
\ifx \shownote     \undefined \def \shownote      #1{#1}          \fi
\ifx \showarticletitle \undefined \def \showarticletitle #1{#1}   \fi
\ifx \showURL      \undefined \def \showURL       {\relax}        \fi
\providecommand\bibfield[2]{#2}
\providecommand\bibinfo[2]{#2}
\providecommand\natexlab[1]{#1}
\providecommand\showeprint[2][]{arXiv:#2}

\bibitem[Chen et~al\mbox{.}(2025a)]%
        {chen2025iterqr}
\bibfield{author}{\bibinfo{person}{Shangyu Chen}, \bibinfo{person}{Xinyu Jia}, \bibinfo{person}{Yingfei Zhang}, \bibinfo{person}{Shuai Zhang}, \bibinfo{person}{Xiang Li}, {and} \bibinfo{person}{Wei Lin}.} \bibinfo{year}{2025}\natexlab{a}.
\newblock \showarticletitle{IterQR: An Iterative Framework for LLM-based Query Rewrite in e-Commercial Search System}.
\newblock \bibinfo{journal}{\emph{arXiv preprint arXiv:2504.05309}} (\bibinfo{year}{2025}).
\newblock


\bibitem[Chen et~al\mbox{.}(2025b)]%
        {chen2025improving}
\bibfield{author}{\bibinfo{person}{Yiqun Chen}, \bibinfo{person}{Lingyong Yan}, \bibinfo{person}{Weiwei Sun}, \bibinfo{person}{Xinyu Ma}, \bibinfo{person}{Yi Zhang}, \bibinfo{person}{Shuaiqiang Wang}, \bibinfo{person}{Dawei Yin}, \bibinfo{person}{Yiming Yang}, {and} \bibinfo{person}{Jiaxin Mao}.} \bibinfo{year}{2025}\natexlab{b}.
\newblock \showarticletitle{Improving retrieval-augmented generation through multi-agent reinforcement learning}.
\newblock \bibinfo{journal}{\emph{arXiv preprint arXiv:2501.15228}} (\bibinfo{year}{2025}).
\newblock


\bibitem[Cuconasu et~al\mbox{.}(2024)]%
        {cuconasu2024power}
\bibfield{author}{\bibinfo{person}{Florin Cuconasu}, \bibinfo{person}{Giovanni Trappolini}, \bibinfo{person}{Federico Siciliano}, \bibinfo{person}{Simone Filice}, \bibinfo{person}{Cesare Campagnano}, \bibinfo{person}{Yoelle Maarek}, \bibinfo{person}{Nicola Tonellotto}, {and} \bibinfo{person}{Fabrizio Silvestri}.} \bibinfo{year}{2024}\natexlab{}.
\newblock \showarticletitle{The power of noise: Redefining retrieval for rag systems}. In \bibinfo{booktitle}{\emph{Proceedings of the 47th International ACM SIGIR Conference on Research and Development in Information Retrieval}}. \bibinfo{pages}{719--729}.
\newblock


\bibitem[Dai et~al\mbox{.}(2024)]%
        {dai2024enhancing}
\bibfield{author}{\bibinfo{person}{Aijun Dai}, \bibinfo{person}{Zhenyu Zhu}, \bibinfo{person}{Haiqing Hu}, \bibinfo{person}{Guoyu Tang}, \bibinfo{person}{Lin Liu}, {and} \bibinfo{person}{Sulong Xu}.} \bibinfo{year}{2024}\natexlab{}.
\newblock \showarticletitle{Enhancing E-Commerce Query Rewriting: A Large Language Model Approach with Domain-Specific Pre-Training and Reinforcement Learning}. In \bibinfo{booktitle}{\emph{Proceedings of the 33rd ACM International Conference on Information and Knowledge Management}}. \bibinfo{pages}{4439--4445}.
\newblock


\bibitem[Fang et~al\mbox{.}(2024)]%
        {fang2024enhancing}
\bibfield{author}{\bibinfo{person}{Feiteng Fang}, \bibinfo{person}{Yuelin Bai}, \bibinfo{person}{Shiwen Ni}, \bibinfo{person}{Min Yang}, \bibinfo{person}{Xiaojun Chen}, {and} \bibinfo{person}{Ruifeng Xu}.} \bibinfo{year}{2024}\natexlab{}.
\newblock \showarticletitle{Enhancing noise robustness of retrieval-augmented language models with adaptive adversarial training}.
\newblock \bibinfo{journal}{\emph{arXiv preprint arXiv:2405.20978}} (\bibinfo{year}{2024}).
\newblock


\bibitem[Feng et~al\mbox{.}(2025)]%
        {feng2025complicated}
\bibfield{author}{\bibinfo{person}{Yunling Feng}, \bibinfo{person}{Gui Ling}, \bibinfo{person}{Yue Jiang}, \bibinfo{person}{Jianfeng Huang}, \bibinfo{person}{Dan Ou}, \bibinfo{person}{Qingwen Liu}, \bibinfo{person}{Fuyu Lv}, {and} \bibinfo{person}{Yajing Xu}.} \bibinfo{year}{2025}\natexlab{}.
\newblock \showarticletitle{Complicated Semantic Alignment for Long-Tail Query Rewriting in Taobao Search Based on Large Language Model}. In \bibinfo{booktitle}{\emph{Proceedings of the 31st ACM SIGKDD Conference on Knowledge Discovery and Data Mining V. 2}}. \bibinfo{pages}{4435--4446}.
\newblock


\bibitem[Gamzu et~al\mbox{.}(2020)]%
        {gamzu2020query}
\bibfield{author}{\bibinfo{person}{Iftah Gamzu}, \bibinfo{person}{Marina Haikin}, {and} \bibinfo{person}{Nissim Halabi}.} \bibinfo{year}{2020}\natexlab{}.
\newblock \showarticletitle{Query rewriting for voice shopping null queries}. In \bibinfo{booktitle}{\emph{Proceedings of the 43rd International ACM SIGIR Conference on Research and Development in Information Retrieval}}. \bibinfo{pages}{1369--1378}.
\newblock


\bibitem[Gao et~al\mbox{.}(2024)]%
        {gao2024smartrag}
\bibfield{author}{\bibinfo{person}{Jingsheng Gao}, \bibinfo{person}{Linxu Li}, \bibinfo{person}{Weiyuan Li}, \bibinfo{person}{Yuzhuo Fu}, {and} \bibinfo{person}{Bin Dai}.} \bibinfo{year}{2024}\natexlab{}.
\newblock \showarticletitle{Smartrag: Jointly learn rag-related tasks from the environment feedback}.
\newblock \bibinfo{journal}{\emph{arXiv preprint arXiv:2410.18141}} (\bibinfo{year}{2024}).
\newblock


\bibitem[Gao et~al\mbox{.}(2023)]%
        {gao2023retrieval}
\bibfield{author}{\bibinfo{person}{Yunfan Gao}, \bibinfo{person}{Yun Xiong}, \bibinfo{person}{Xinyu Gao}, \bibinfo{person}{Kangxiang Jia}, \bibinfo{person}{Jinliu Pan}, \bibinfo{person}{Yuxi Bi}, \bibinfo{person}{Yixin Dai}, \bibinfo{person}{Jiawei Sun}, \bibinfo{person}{Haofen Wang}, {and} \bibinfo{person}{Haofen Wang}.} \bibinfo{year}{2023}\natexlab{}.
\newblock \showarticletitle{Retrieval-augmented generation for large language models: A survey}.
\newblock \bibinfo{journal}{\emph{arXiv preprint arXiv:2312.10997}} \bibinfo{volume}{2}, \bibinfo{number}{1} (\bibinfo{year}{2023}).
\newblock


\bibitem[Gong et~al\mbox{.}(2024)]%
        {gong2024cosearchagent}
\bibfield{author}{\bibinfo{person}{Peiyuan Gong}, \bibinfo{person}{Jiamian Li}, {and} \bibinfo{person}{Jiaxin Mao}.} \bibinfo{year}{2024}\natexlab{}.
\newblock \showarticletitle{Cosearchagent: a lightweight collaborative search agent with large language models}. In \bibinfo{booktitle}{\emph{Proceedings of the 47th International ACM SIGIR Conference on Research and Development in Information Retrieval}}. \bibinfo{pages}{2729--2733}.
\newblock


\bibitem[Guu et~al\mbox{.}(2020)]%
        {guu2020retrieval}
\bibfield{author}{\bibinfo{person}{Kelvin Guu}, \bibinfo{person}{Kenton Lee}, \bibinfo{person}{Zora Tung}, \bibinfo{person}{Panupong Pasupat}, {and} \bibinfo{person}{Mingwei Chang}.} \bibinfo{year}{2020}\natexlab{}.
\newblock \showarticletitle{Retrieval augmented language model pre-training}. In \bibinfo{booktitle}{\emph{International conference on machine learning}}. PMLR, \bibinfo{pages}{3929--3938}.
\newblock


\bibitem[Jiang et~al\mbox{.}(2023)]%
        {jiang2023active}
\bibfield{author}{\bibinfo{person}{Zhengbao Jiang}, \bibinfo{person}{Frank~F Xu}, \bibinfo{person}{Luyu Gao}, \bibinfo{person}{Zhiqing Sun}, \bibinfo{person}{Qian Liu}, \bibinfo{person}{Jane Dwivedi-Yu}, \bibinfo{person}{Yiming Yang}, \bibinfo{person}{Jamie Callan}, {and} \bibinfo{person}{Graham Neubig}.} \bibinfo{year}{2023}\natexlab{}.
\newblock \showarticletitle{Active retrieval augmented generation}. In \bibinfo{booktitle}{\emph{Proceedings of the 2023 Conference on Empirical Methods in Natural Language Processing}}. \bibinfo{pages}{7969--7992}.
\newblock


\bibitem[Kakkar et~al\mbox{.}(2023)]%
        {kakkar2023search}
\bibfield{author}{\bibinfo{person}{Vishal Kakkar}, \bibinfo{person}{Chinmay Sharma}, \bibinfo{person}{Madhura Pande}, {and} \bibinfo{person}{Surender Kumar}.} \bibinfo{year}{2023}\natexlab{}.
\newblock \showarticletitle{Search query spell correction with weak supervision in E-commerce}. In \bibinfo{booktitle}{\emph{Proceedings of the 61st Annual Meeting of the Association for Computational Linguistics (Volume 5: Industry Track)}}. \bibinfo{pages}{687--694}.
\newblock


\bibitem[Li et~al\mbox{.}(2022b)]%
        {li2022survey}
\bibfield{author}{\bibinfo{person}{Huayang Li}, \bibinfo{person}{Yixuan Su}, \bibinfo{person}{Deng Cai}, \bibinfo{person}{Yan Wang}, {and} \bibinfo{person}{Lemao Liu}.} \bibinfo{year}{2022}\natexlab{b}.
\newblock \showarticletitle{A survey on retrieval-augmented text generation}.
\newblock \bibinfo{journal}{\emph{arXiv preprint arXiv:2202.01110}} (\bibinfo{year}{2022}).
\newblock


\bibitem[Li et~al\mbox{.}(2022a)]%
        {li2022query}
\bibfield{author}{\bibinfo{person}{Sen Li}, \bibinfo{person}{Fuyu Lv}, \bibinfo{person}{Taiwei Jin}, \bibinfo{person}{Guiyang Li}, \bibinfo{person}{Yukun Zheng}, \bibinfo{person}{Tao Zhuang}, \bibinfo{person}{Qingwen Liu}, \bibinfo{person}{Xiaoyi Zeng}, \bibinfo{person}{James Kwok}, {and} \bibinfo{person}{Qianli Ma}.} \bibinfo{year}{2022}\natexlab{a}.
\newblock \showarticletitle{Query rewriting in taobao search}. In \bibinfo{booktitle}{\emph{Proceedings of the 31st ACM International Conference on Information \& Knowledge Management}}. \bibinfo{pages}{3262--3271}.
\newblock


\bibitem[Liu et~al\mbox{.}(2020)]%
        {liu2020query}
\bibfield{author}{\bibinfo{person}{Hui Liu}, \bibinfo{person}{Dawei Yin}, {and} \bibinfo{person}{Jiliang Tang}.} \bibinfo{year}{2020}\natexlab{}.
\newblock \showarticletitle{Query rewriting}.
\newblock In \bibinfo{booktitle}{\emph{Query Understanding for Search Engines}}. \bibinfo{publisher}{Springer}, \bibinfo{pages}{129--144}.
\newblock


\bibitem[Mackie et~al\mbox{.}(2023)]%
        {mackie2023generative}
\bibfield{author}{\bibinfo{person}{Iain Mackie}, \bibinfo{person}{Shubham Chatterjee}, {and} \bibinfo{person}{Jeffrey Dalton}.} \bibinfo{year}{2023}\natexlab{}.
\newblock \showarticletitle{Generative relevance feedback with large language models}. In \bibinfo{booktitle}{\emph{Proceedings of the 46th international ACM SIGIR conference on research and development in information retrieval}}. \bibinfo{pages}{2026--2031}.
\newblock


\bibitem[Mei et~al\mbox{.}(2025)]%
        {mei2025survey}
\bibfield{author}{\bibinfo{person}{Lang Mei}, \bibinfo{person}{Siyu Mo}, \bibinfo{person}{Zhihan Yang}, {and} \bibinfo{person}{Chong Chen}.} \bibinfo{year}{2025}\natexlab{}.
\newblock \showarticletitle{A survey of multimodal retrieval-augmented generation}.
\newblock \bibinfo{journal}{\emph{arXiv preprint arXiv:2504.08748}} (\bibinfo{year}{2025}).
\newblock


\bibitem[Naseri et~al\mbox{.}(2021)]%
        {naseri2021ceqe}
\bibfield{author}{\bibinfo{person}{Shahrzad Naseri}, \bibinfo{person}{Jeffrey Dalton}, \bibinfo{person}{Andrew Yates}, {and} \bibinfo{person}{James Allan}.} \bibinfo{year}{2021}\natexlab{}.
\newblock \showarticletitle{Ceqe: Contextualized embeddings for query expansion}. In \bibinfo{booktitle}{\emph{European conference on information retrieval}}. Springer, \bibinfo{pages}{467--482}.
\newblock


\bibitem[Peng et~al\mbox{.}(2024)]%
        {peng2024large}
\bibfield{author}{\bibinfo{person}{Wenjun Peng}, \bibinfo{person}{Guiyang Li}, \bibinfo{person}{Yue Jiang}, \bibinfo{person}{Zilong Wang}, \bibinfo{person}{Dan Ou}, \bibinfo{person}{Xiaoyi Zeng}, \bibinfo{person}{Derong Xu}, \bibinfo{person}{Tong Xu}, {and} \bibinfo{person}{Enhong Chen}.} \bibinfo{year}{2024}\natexlab{}.
\newblock \showarticletitle{Large language model based long-tail query rewriting in taobao search}. In \bibinfo{booktitle}{\emph{Companion Proceedings of the ACM Web Conference 2024}}. \bibinfo{pages}{20--28}.
\newblock


\bibitem[Qiu et~al\mbox{.}(2021)]%
        {qiu2021query}
\bibfield{author}{\bibinfo{person}{Yiming Qiu}, \bibinfo{person}{Kang Zhang}, \bibinfo{person}{Han Zhang}, \bibinfo{person}{Songlin Wang}, \bibinfo{person}{Sulong Xu}, \bibinfo{person}{Yun Xiao}, \bibinfo{person}{Bo Long}, {and} \bibinfo{person}{Wen-Yun Yang}.} \bibinfo{year}{2021}\natexlab{}.
\newblock \showarticletitle{Query rewriting via cycle-consistent translation for e-commerce search}. In \bibinfo{booktitle}{\emph{2021 IEEE 37th International Conference on Data Engineering (ICDE)}}. IEEE, \bibinfo{pages}{2435--2446}.
\newblock


\bibitem[Shao et~al\mbox{.}(2024)]%
        {shao2024deepseekmath}
\bibfield{author}{\bibinfo{person}{Zhihong Shao}, \bibinfo{person}{Peiyi Wang}, \bibinfo{person}{Qihao Zhu}, \bibinfo{person}{Runxin Xu}, \bibinfo{person}{Junxiao Song}, \bibinfo{person}{Xiao Bi}, \bibinfo{person}{Haowei Zhang}, \bibinfo{person}{Mingchuan Zhang}, \bibinfo{person}{YK Li}, \bibinfo{person}{Yang Wu}, {et~al\mbox{.}}} \bibinfo{year}{2024}\natexlab{}.
\newblock \showarticletitle{Deepseekmath: Pushing the limits of mathematical reasoning in open language models}.
\newblock \bibinfo{journal}{\emph{arXiv preprint arXiv:2402.03300}} (\bibinfo{year}{2024}).
\newblock


\bibitem[Si et~al\mbox{.}(2023)]%
        {si2023search}
\bibfield{author}{\bibinfo{person}{Zihua Si}, \bibinfo{person}{Zhongxiang Sun}, \bibinfo{person}{Xiao Zhang}, \bibinfo{person}{Jun Xu}, \bibinfo{person}{Xiaoxue Zang}, \bibinfo{person}{Yang Song}, \bibinfo{person}{Kun Gai}, {and} \bibinfo{person}{Ji-Rong Wen}.} \bibinfo{year}{2023}\natexlab{}.
\newblock \showarticletitle{When search meets recommendation: Learning disentangled search representation for recommendation}. In \bibinfo{booktitle}{\emph{Proceedings of the 46th international ACM SIGIR conference on research and development in information retrieval}}. \bibinfo{pages}{1313--1323}.
\newblock


\bibitem[Song and Zheng(2024)]%
        {song2024survey}
\bibfield{author}{\bibinfo{person}{Mingyang Song} {and} \bibinfo{person}{Mao Zheng}.} \bibinfo{year}{2024}\natexlab{}.
\newblock \showarticletitle{A Survey of Query Optimization in Large Language Models}.
\newblock \bibinfo{journal}{\emph{arXiv preprint arXiv:2412.17558}} (\bibinfo{year}{2024}).
\newblock


\bibitem[Sun et~al\mbox{.}(2024)]%
        {sun2024rethinking}
\bibfield{author}{\bibinfo{person}{Hao Sun}, \bibinfo{person}{Yunyi Shen}, {and} \bibinfo{person}{Jean-Francois Ton}.} \bibinfo{year}{2024}\natexlab{}.
\newblock \showarticletitle{Rethinking bradley-terry models in preference-based reward modeling: Foundations, theory, and alternatives}.
\newblock \bibinfo{journal}{\emph{arXiv preprint arXiv:2411.04991}} (\bibinfo{year}{2024}).
\newblock


\bibitem[Sun et~al\mbox{.}(2023)]%
        {sun2023kuaisar}
\bibfield{author}{\bibinfo{person}{Zhongxiang Sun}, \bibinfo{person}{Zihua Si}, \bibinfo{person}{Xiaoxue Zang}, \bibinfo{person}{Dewei Leng}, \bibinfo{person}{Yanan Niu}, \bibinfo{person}{Yang Song}, \bibinfo{person}{Xiao Zhang}, {and} \bibinfo{person}{Jun Xu}.} \bibinfo{year}{2023}\natexlab{}.
\newblock \showarticletitle{KuaiSar: A unified search and recommendation dataset}. In \bibinfo{booktitle}{\emph{Proceedings of the 32nd ACM international conference on information and knowledge management}}. \bibinfo{pages}{5407--5411}.
\newblock


\bibitem[Wang et~al\mbox{.}(2023a)]%
        {wang2023learning}
\bibfield{author}{\bibinfo{person}{Binbin Wang}, \bibinfo{person}{Mingming Li}, \bibinfo{person}{Zhixiong Zeng}, \bibinfo{person}{Jingwei Zhuo}, \bibinfo{person}{Songlin Wang}, \bibinfo{person}{Sulong Xu}, \bibinfo{person}{Bo Long}, {and} \bibinfo{person}{Weipeng Yan}.} \bibinfo{year}{2023}\natexlab{a}.
\newblock \showarticletitle{Learning multi-stage multi-grained semantic embeddings for e-commerce search}. In \bibinfo{booktitle}{\emph{Companion Proceedings of the ACM Web Conference 2023}}. \bibinfo{pages}{411--415}.
\newblock


\bibitem[Wang et~al\mbox{.}(2023b)]%
        {wang2023query2doc}
\bibfield{author}{\bibinfo{person}{Liang Wang}, \bibinfo{person}{Nan Yang}, {and} \bibinfo{person}{Furu Wei}.} \bibinfo{year}{2023}\natexlab{b}.
\newblock \showarticletitle{Query2doc: Query expansion with large language models}.
\newblock \bibinfo{journal}{\emph{arXiv preprint arXiv:2303.07678}} (\bibinfo{year}{2023}).
\newblock


\bibitem[Xiao et~al\mbox{.}(2019)]%
        {xiao2019weakly}
\bibfield{author}{\bibinfo{person}{Rong Xiao}, \bibinfo{person}{Jianhui Ji}, \bibinfo{person}{Baoliang Cui}, \bibinfo{person}{Haihong Tang}, \bibinfo{person}{Wenwu Ou}, \bibinfo{person}{Yanghua Xiao}, \bibinfo{person}{Jiwei Tan}, {and} \bibinfo{person}{Xuan Ju}.} \bibinfo{year}{2019}\natexlab{}.
\newblock \showarticletitle{Weakly supervised co-training of query rewriting andsemantic matching for e-commerce}. In \bibinfo{booktitle}{\emph{Proceedings of the twelfth ACM international conference on web search and data mining}}. \bibinfo{pages}{402--410}.
\newblock


\bibitem[Yan et~al\mbox{.}(2024)]%
        {yan2024corrective}
\bibfield{author}{\bibinfo{person}{Shi-Qi Yan}, \bibinfo{person}{Jia-Chen Gu}, \bibinfo{person}{Yun Zhu}, {and} \bibinfo{person}{Zhen-Hua Ling}.} \bibinfo{year}{2024}\natexlab{}.
\newblock \showarticletitle{Corrective retrieval augmented generation}.
\newblock  (\bibinfo{year}{2024}).
\newblock


\bibitem[Ye et~al\mbox{.}(2023)]%
        {ye2023improving}
\bibfield{author}{\bibinfo{person}{Dezhi Ye}, \bibinfo{person}{Bowen Tian}, \bibinfo{person}{Jiabin Fan}, \bibinfo{person}{Jie Liu}, \bibinfo{person}{Tianhua Zhou}, \bibinfo{person}{Xiang Chen}, \bibinfo{person}{Mingming Li}, {and} \bibinfo{person}{Jin Ma}.} \bibinfo{year}{2023}\natexlab{}.
\newblock \showarticletitle{Improving query correction using pre-train language model in search engines}. In \bibinfo{booktitle}{\emph{Proceedings of the 32nd ACM International Conference on Information and Knowledge Management}}. \bibinfo{pages}{2999--3008}.
\newblock


\bibitem[Zuo et~al\mbox{.}(2025)]%
        {zuo2025value}
\bibfield{author}{\bibinfo{person}{Boyang Zuo}, \bibinfo{person}{Xiao Zhang}, \bibinfo{person}{Feng Li}, \bibinfo{person}{Pengjie Wang}, \bibinfo{person}{Jian Xu}, {and} \bibinfo{person}{Bo Zheng}.} \bibinfo{year}{2025}\natexlab{}.
\newblock \showarticletitle{VALUE: Value-Aware Large Language Model for Query Rewriting via Weighted Trie in Sponsored Search}.
\newblock \bibinfo{journal}{\emph{arXiv preprint arXiv:2504.05321}} (\bibinfo{year}{2025}).
\newblock


\bibitem[Zuo et~al\mbox{.}(2022)]%
        {zuo2022context}
\bibfield{author}{\bibinfo{person}{Simiao Zuo}, \bibinfo{person}{Qingyu Yin}, \bibinfo{person}{Haoming Jiang}, \bibinfo{person}{Shaohui Xi}, \bibinfo{person}{Bing Yin}, \bibinfo{person}{Chao Zhang}, {and} \bibinfo{person}{Tuo Zhao}.} \bibinfo{year}{2022}\natexlab{}.
\newblock \showarticletitle{Context-Aware Query Rewriting for Improving Users' Search Experience on E-commerce Websites}.
\newblock \bibinfo{journal}{\emph{arXiv preprint arXiv:2209.07584}} (\bibinfo{year}{2022}).
\newblock


\end{thebibliography}

\newpage

\appendix

\section{Detailed Prompts}
\label{prompts}
We outline the prompt details for implementing the CardRewriter rewriting workflow to facilitate the reproducibility of our work.
\begin{itemize}
    \item Card Generation. Summarizing the multi-source knowledge collected from the short-video platform into an informative and concise knowledge card that remains relevant to the original query, as shown in Table \ref{prompt1}.
    \item Card-based Rewriting. leveraging the generated knowledge card to better capture and understand the search context, thereby producing higher-quality query rewrites, as illustrated in Table \ref{prompt2}.
\end{itemize}

\section{Offline Metrics}
\label{appendix_metrics}
We provide a detailed description of how to implement the three metrics: Rel, Increment, and Hitrate@$K$.
\begin{itemize}
\item Rel (Relevance): evaluates the semantic quality of model outputs.
For card generation, it measures how relevant the generated card is to the original query (QC-Rel).
For card-based rewriting, it assesses: (I) the semantic relevance between the rewritten and the original queries., and (II) Whether the rewriting effectively integrates the content of the knowledge card (QR-Rel).
We utilize Qwen3-235B-A22B to provide binary judgments (1 = success, 0 = failure) for each instance.
\item Increment (Retrieval Expansion): Quantifies the model’s ability to expand retrieval coverage by comparing recall obtained from the original query with that from its rewritten counterpart.
let $V_X$ and $V_Y$ denote the video lists retrieved for the original query $x$ and the rewritten query $y$, respectively. 
Then, Increment is defined as:
\begin{equation}
\text{Increment} = \frac{|V_X \cup V_Y| - |V_X|}{|V_X|}.
\end{equation}     
\item Hitrate@$K$ (User Satisfaction): Assesses whether the rewritten query can retrieve videos that align with user intent. Given an input query $x$, we construct a ground-truth set $\mathcal{E}$ consisting of videos that users either clicked on or watched for an extended period after reformulating $x$ within a one-week window. Hitrate@$K$ is then computed as:
 \begin{equation} \text{Hitrate@K} = \begin{cases} 1, & \text{if } \exists v_i \in \mathcal{E},\ i \leq K \\ 0, & \text{otherwise} \end{cases} \end{equation}
where $v_i$ denotes a video retrieved for query $x$.
\end{itemize}

\begin{table*}[t]
    \centering
        \caption{Card Generation Prompt. Due to the excessive length of the prompt, we have made some omissions in the section where video information is inserted.}
    \begin{tabularx}{\linewidth}{X}
        \toprule
        \textbf{Role} \\
        You are a short video search expert. Your core task is to analyze and understand the user’s [original query], and—by combining it with the retrieved short video information and related general information—generate a [demand description analysis] within 200 characters.
        The goal is to help the short video search platform better understand the user’s true intent, so it can use the [original query] together with the [demand description analysis] to rewrite an improved [rewritten query], thereby helping users find more relevant and effective short videos.\\
        \textbf{Task Requirements} \\
        1. Short video information includes: short video title, cover OCR text, author name, background music name, and key frames.
        General information refers to open-domain data related to the original query.\\
        2. You need to use this information to determine whether the searched videos are relevant to the [original query].
        Since user queries may contain ambiguous expressions, you should use the relevant information comprehensively to clarify the true referent (e.g., a specific streamer, short drama title, or game name) and specify it in the [demand description analysis].\\
        3. The generated [demand description analysis] must be concise and effective, helping the platform accurately understand the user’s search intent.
        Do not include content irrelevant to entities or attributes in the [original query].
        Do not include phrases like “based on video information” or “according to search results,” and do not give instructions on how to rewrite the original query.
        Only provide an analysis of the intent behind the original query based on the retrieved videos.\\
        4. Output in JSON format. The desc field represents your [demand description analysis].
        The final output must contain only one JSON result, with no extra characters.\\
        \textbf{Role} \\
        –– Original Query:\\
        {{Query}}\\
        –– Short Video Information: \\
        <Video 1> \\
        –– Title: {{title}} \\
        –– Cover OCR Text: {{OCR}} \\
        –– Author Name: {{author name}} \\
        –– Background Music Name: {{BGE name}} \\
        –– Key Frames: {{frame1}}; {{frame2}}; {{frame3}} \\
        </Video 1> \\
        <Video 2> \\
        … \\
        </Video 2> \\
        <Video 3> \\
        … \\
        </Video 3> \\
        –– General Information: \\
        {{Open-domain Knowledge}} \\
        \bottomrule
    \end{tabularx}
    \label{prompt1}
\end{table*}

\begin{table*}[t]
    \centering
        \caption{Card-based Rewriting Prompt}
    \begin{tabularx}{\linewidth}{X}
        \toprule
        \textbf{Role} \\
        You are an expert specializing in optimizing user search queries for short video platforms. Your core task is to understand the user's [original search query] and rewrite it into a [rewritten query] that is more easily understood by short video search engines. The information provided to you includes the [original search query] and the [original search query needs analysis] provided by search experts. You can combine this needs analysis to understand the [original search query] and make targeted rewrites. While maintaining semantic consistency, your [rewritten query] can retrieve additional relevant videos that the [original search query] could not.\\
        \textbf{Task Requirements} \\
        \textbf{====Query Analysis Requirements====} \\
        1. Short video search queries have the following characteristics: **strong domain knowledge, colloquial and concise descriptions, vague or broad intent, and a high number of typos**. These characteristics must be considered when analyzing user search needs.\\
        2. When analyzing video search needs in user queries, consider the core intent, key entities, and attribute constraints. The necessary attribute constraints that appear in each query should be considered. You can refer to the content in the [Original Search Query Requirement Analysis] to help you correctly analyze and understand the [Original Search Query].\\
        \textbf{====Query Rewriting Requirements====}\\
        1. Based on your analysis, rewrite the original search query from the perspective of a video search expert. **The user's core intent and original needs must not change before and after the rewriting**. Avoid changing the scope of the user's original needs, such as removing key attributes or adding irrelevant attribute restrictions.\\
        2. The rewritten query should be as different as possible from the original query, such as by changing the wording and description to increase the number of videos retrieved and prevent excessive duplication of results between the rewritten and original queries.\\
        3. The rewritten query should be concise and clear. Replace complex user descriptions with concise synonyms. Whenever possible, rewrite long queries to shorten them while preserving the semantics, rather than lengthening short queries. Avoid adding meaningless suffixes such as "xxx video" or "xxx related videos" as rewrite suffixes, and avoid translating English words in the query. If the user's query contains irrelevant symbols such as \#, remove them. \\
        4. For long and difficult user queries, use keywords in parallel as much as possible, rather than piling them into long declarative sentences.
        5. Domain terms that are still unclear after analyzing the [Requirements Analysis] should be output as is. Do not attempt to guess their meaning. If the user's original search query contains usernames, short drama titles, or other proper nouns related to possible short video scenes, do not modify them. \\
        \textbf{====Output Requirements====}\\
        1. Output in JSON format. The RewriteQuery field represents [Rewritten Query].\\
        2. Do not output any irrelevant symbols or descriptions other than the output JSON.\\
        \textbf{Task Start} \\
        Original Search Query: \{\{Query\}\} \\
        Original Search Query Requirements Analysis: \{\{Card\}\} \\
        \bottomrule
    \end{tabularx}
    \label{prompt2}
\end{table*}

\section{Implementation Details}
\label{details}
The training of CardRewriter proceeds through three stages: Supervised Fine-Tuning, Reward Modeling that captures system preferences, and Objective Alignment. 
All experiments were conducted utilizing two 8-GPU H800 nodes.

\textbf{Supervised Fine-Tuning} We employ Qwen2.5-VL-72B to generate knowledge cards for constructing the initial dataset in the card generation task, and Qwen3-235B-A22B to produce rewritten queries for building the initial dataset in the query rewriting task. 
For card generation, we fine-tune the Qwen2.5-VL-7B-Instruct model with a learning rate of $1 \times 10^{-5}$ using the AdamW optimizer $(\beta_{1}=0.9, \beta_{2}=0.999)$ and a weight decay of $0.01$, running training for one epoch with a per-device batch size of $16$ and gradient accumulation of $4$ steps.
The vision tower and multi-modal projector remain frozen while the language model parameters are fully fine-tuned. 
We employ DeepSpeed ZeRO-3 for distributed training efficiency and conduct all computations in \texttt{bfloat16} precision. 
For query rewriting, we fine-tune the Qwen3-8B model with a learning rate of $1 \times 10^{-5}$ using the AdamW optimizer $(\beta_{1}=0.9, \beta_{2}=0.999)$ and a weight decay of $0.01$, running training for two epochs with a per-device batch size of $16$ and gradient accumulation of $4$ steps. 
We enable FlashAttention-2 and Liger kernels to accelerate computation and perform all operations in \texttt{bfloat16} precision. A cosine learning rate scheduler with a warmup ratio of $0.1$ adjusts the learning rate, and DeepSpeed ZeRO-3 is used for distributed optimization.

\textbf{Reward Modeling} 
We fine-tune the Qwen3-8B model (Classification version) with a learning rate of $1 \times 10^{-5}$ using the AdamW optimizer $(\beta_{1}=0.9, \beta_{2}=0.999)$ and a weight decay of $0.01$. 
Training runs for two epochs with a per-device batch size of $16$ and a gradient accumulation step of $4$. 
All computations are performed in \texttt{bfloat16} precision to improve training efficiency, and the maximum sequence length is set to $512$ tokens.

\textbf{Objective Alignment} 
For the card generation task, we adopt the SFT-based card generation model combined with GRPO advantage estimation. The hyperparameters are configured as follows: training batch size of $512$, maximum prompt length of $10{,}240$ tokens, and maximum response length of $512$ tokens. The actor model is optimized with a learning rate of $1\times10^{-6}$. The KL loss coefficient is set to $0.01$ using the low-variance KL formulation, and entropy regularization is disabled. Rollouts are generated via the vLLM engine with a tensor model parallel size of $2$, GPU memory utilization of $0.6$, chunked prefill enabled, and a rollout sample size of $n=8$. The training process consists of a single epoch.
For the query rewriting task, we employ an SFT-based query rewriting model with the following hyperparameters: training batch size of $128$, maximum prompt length of $2{,}048$ tokens, and maximum response length of $128$ tokens. The actor model is optimized with a learning rate of $1\times10^{-6}$, trained using dynamic batch sizes under the FSDP2 distributed strategy. The KL loss coefficient is set to $0.001$ with the low-variance KL formulation, and entropy regularization is disabled. Rollouts are generated via the vLLM engine with a rollout sample size of $n=8$, sampling temperature $T=1.0$, top-$p=1.0$, and top-$k=-1$. The training process also consists of a single epoch.

\end{document}